\documentclass[10pt]{amsart}
\usepackage{comment,balance,pdfpages}
\usepackage{soul}
\usepackage{colortbl}
\usepackage{dsfont}
\usepackage{amsthm}
\usepackage{stmaryrd}
\usepackage{amsmath,amssymb,amsfonts,bbm,multicol,caption}
\usepackage{graphicx}
\usepackage{tcolorbox}
\usepackage{xcolor,braket}
\usepackage{multirow,flushend}
\usepackage{array} 
\usepackage{longtable}
\usepackage{url}
\usepackage{tabularx} 
\usepackage{algorithm,algorithmic}
\usepackage{float}
\usepackage{tabularx}
\usepackage{booktabs}

\usepackage{tikz}
\usetikzlibrary{positioning, shapes, arrows}

\newcommand{\E}{\mathbb{E}}
\renewcommand{\P}{\mathbb{P}}

\usepackage{mathrsfs}
\usepackage[scr=rsfs,cal=boondox]{mathalfa}

\PassOptionsToPackage{hyphens}{url}
\usepackage{xurl}
\usepackage{acronym}
\usepackage[acronym]{glossaries}
\usepackage{glossary-mcols}

\begin{document}
\title{An efficient Progressive Swapping to the Middle distribution protocol adapted to imperfect quantum memories in quantum networks}
	\author{Claire Mesny}
   \address[Claire Mesny]{Orange Innovation}
\curraddr{2, Avenue Pierre Marzin, F-22300 Lannion}
\email{claire.mesny@orange.com} 
  \thanks{Presented at 2026 EuCNC \& 6G Summit, 2-5 June, M\'alaga, Spain}

\author{Fabrice Guillemin}
\address[Fabrice Guillemin]{Orange Innovation}
\curraddr{2, Avenue Pierre Marzin, F-22300 Lannion}
\email{fabrice.guillemin@orange.com}

        \author{Claire Goursaud}
   \address[Claire Goursaud]{INSA Lyon}
\curraddr{Institut National des Sciences Appliquées de Lyon, F-69000 Lyon}
\email{claire.goursaud@insa-lyon.fr}     


\date{\today}

\begin{abstract}
The distribution of entangled pairs of photons on the links composing a quantum network, combined with Bell state measurements and teleportation, is the basic apparatus to transfer quantum bits (qubits) over long distances. Entanglement distribution establishes an end-to-end entangled pair while consuming intermediate pairs on links and holding them for a certain time period. The technical literature identifies two main kinds of protocols, parallel and sequential ones, the latter having an advantage in resource consumption over the former. In this paper, we introduce an efficient swapping protocol called Progressive Swapping to the Middle (PSM) as it combines the existing Progressive Swapping (PS) protocol from both extremities of a path that meet in the middle  where the received pairs are swapped. We compare PSM with two parallel protocols and PS; in our evaluation, we take into account imperfect memories and fidelity degradation. We demonstrate that PSM yields a much better link probability than PS while keeping a reasonable link fidelity, and shows an advantage in resource consumption over other protocols.
\end{abstract}

\maketitle

\section{Introduction}
Quantum computers and quantum sensors are developing very rapidly and will likely be very soon  commercialized, in addition to Quantum Key Distribution services already available on the market. Connecting machines performing quantum computations will certainly become a need in the next few years in order to fully benefit from the power of distributed quantum computing. This urges the development of a quantum Internet. The major challenge when designing a quantum Internet is to transfer quantum bits (qubits) over long distances as their interaction with the transmission degrades quantum information. The basic building block for transferring quantum bits is to establish an entangled pair between quantum devices, which is subsequently used for the teleportation of an information qubit between the devices. 

To overcome the problem of  distance, the technique so far used in the technical literature is based on distribution of entangled pairs on the links composing a path. An end-to-end entangled pair is then created by using Bell state measurements. The measurements results are finally transmitted by means of classical communications which are eventually used at the end points of the path to create an end-to-end entangled pair, used for teleporting the information qubit. The distribution and swap of intermediate entangled pairs follow a so-called entanglement distribution protocol. The studied protocols can be separated into sequential protocols (SP) \cite{Wang24_SES_noF_nor_mem,Pouryousef2024_asynchro_SES,Hughes:2025_asynchro_SES_<_PES} with swaps occurring hop by hop, and  parallel protocols (PP) \cite{Chakraborty_2020,Duer99,Abane25} with all nodes swapping simultaneously.

PP are usually more efficient in the sense that their success probability is high \cite{Zhonghui25_sendresourcestobottlnecks,Hughes:2025_asynchro_SES_<_PES}. In particular \cite{Zhonghui25_sendresourcestobottlnecks} observes that parallel protocols are faster than sequential ones, but it does not account for the communication time of the swap results. Additionally, \cite{Hughes:2025_asynchro_SES_<_PES} and \cite{Abane25} work with loss-less memories while these protocols heavily rely on quantum memories. To preserve link fidelity, purification  schemes have been considered on these protocols in \cite{Duer99,Hughes:2025_asynchro_SES_<_PES,li22fidel,Zhonghui25_sendresourcestobottlnecks}.

The advantage of sequential protocols (SP) is developed in \cite{Yang25_routing_best_resource_alloc_ses} and \cite{Pouryousef2024_asynchro_SES} with \cite{Yang25_routing_best_resource_alloc_ses} claiming that SP are more adaptive to dynamical resource allocation when it comes to routing strategies. These works do not, however, count the number of consumed resources (generated pairs) by the protocols. The work in \cite{Wang24_SES_noF_nor_mem} develops a Loss-LES protocol taking advantage of loss-flow for resources, using an adapted expected throughput metric and dealing with bottlenecks in multi-user networks. It makes no mention of the link fidelity nor of fiber losses in the link probability. Similarly to \cite{Pouryousef2024_asynchro_SES} it does not include memory in the protocol success probability either. The work in \cite{Zang_2023_F_decoh_mem} studies a nested protocol with both dephasing and depolarizing memories, but it gives no expression for probability and end-to-end fidelity.

We develop in this paper a distribution protocol that mixes PS and SS and we call it Progressive Swapping to the Middle (PSM). We compare it with the usual SP, Progressive swapping (PS), and to Blind Swapping (BS, also known as Swap ASAP) and Heralded Distribution (HD, Unheralded Swapping in \cite{Abane25}); these two latter protocols being PP. BS is the most straightforward PP with good link fidelity, and HD is a PP with high link probability developed in \cite{Abane25}. 

We compute the success probability of these four protocols for imperfect memories and with exponential decay in the optical fiber, which were not considered in \cite{Duer99,Wang24_SES_noF_nor_mem,Pouryousef2024_asynchro_SES,Hughes:2025_asynchro_SES_<_PES,Abane25}. We also assess the total number of generated pairs for each protocol, which depends on the link probability to establish at least one end-to-end pair on a given path (henceforth called link probability). This metric should help with network scaling and be an additional parameter to consider in routing \cite{Lutong24_routing_for_SES}.

We also derive a formula for the link fidelity of each protocol, which is not considered in \cite{Wang24_SES_noF_nor_mem,Yang25_routing_best_resource_alloc_ses}. We model fidelity decoherence due to swap such as in \cite{Chakraborty_2020,Duer99}, and we add another decay due to interactions in imperfect memories and fibers. Contrary to \cite{Pouryousef2024_asynchro_SES} and \cite{Hughes:2025_asynchro_SES_<_PES}, that model dephasing in memory, we consider decoherence in both memory and fiber, as it is the worst case assumption \cite{Zhonghui25_sendresourcestobottlnecks}.

The basic principles of our model are detailed section \ref{part:basic_principles}, followed by the parallel protocols in section \ref{part:parallel_swapping} and sequential protocols section \ref{part:sequential_distribution}. We compare the protocols performances in section \ref{part:performances} before concluding in section \ref{part:conclusion}.





\section{Basic principles}
\label{part:basic_principles}

In order to establish an end-to-end entangled pair between a source and a destination, $n$ pairs of ebits are initially generated at once (also called width in \cite{Dupuy23}). We consider that the sources are perfect and there is no delay between two generated pairs. In parallel protocols, all nodes initially generate $n$ pairs before sending photons to their nearest neighbors. Sequential protocols initially generate $n$ pairs only at the source node. One half of each pair is stored in the local memory of the node and the other half is sent to the next node through the connecting fiber. Sequential protocols then generate and swap a locally decided amount of pairs at the local node until destination is reached.

In this paper, we consider that time is slotted with time slot duration denoted by $\Delta$. This time is greater than the delay it takes for a photon to reach a neighboring node at distance $L$ with speed $c$, so that $\Delta > L/c$. We model a traveling photon through a fiber of loss parameter $\alpha$ (typically $0.2dB/km$ at telecom wavelength) to have an exponential survival probability $e^{-\alpha L/10}.$ 

When stored in memory, the photons have a probability $e^{- t/\tau}$ to be absorbed after a time $t$, in which $\tau$ is the mean qubit life time in the memory. To simplify, we  assume that write-in and read-out operations are perfect, as well as the sources that deterministically generate the expected pairs with fidelity $1$. We  denote by $L_i$ the distance between node $i$ and node $i+1$, and $\tau_i$ the memory life-time of node $i$. There are $l_r$ nodes on the path; the source is node $1$ and the destination is node $l_r$.

A swap operation at each intermediary node can  make one long distance ebit pair out of the  ebits distributed on adjacent links. Optical swaps, using Bell State Measurement and photons, are fundamentally limited by a success probability $p_{swap} = 0.5$. Some set-ups use ancillary qubits and achieve a success probability greater than $0.6$ and theoretically even $0.95$ at the expense of experimental complexity \cite{Kilmer_2019}. In the following, we assume that intermediate nodes can generate as many ebits as necessary in order to concentrate on the performance of the entanglement distribution protocol. For the same reason, we chose $p_{swap} = 0.9$ corresponding to an almost deterministic swap operation. 

Deterministic swap operations are being studied, but the technical difficulty makes these process inefficient \cite{Arenskoetter_deterministicBSM_low_eff,Kamimaki:2022_deterministic_BSM}. They also raise the question of hybrid systems with different types of hardware which induces some more losses.

Time also plays a role in qubit decoherence, so that fidelity decreases from $F(t_0)$ at time $t_0$ to \cite{Pouryousef2024_asynchro_SES}\begin{equation}
F_{\delta}(t, F(t_0)) = \frac{1}{4} + (F(t_0) - \frac{1}{4})e^{-(t-t_0)/\delta}\label{eq:F_decoh}
\end{equation} in which $\delta$ is the mean life time either in a memory or in the fiber. For the fiber we take an equivalent $\delta = 10/(\alpha c)$. 

For all nodes, we consider that the probability for a swap to succeed is $p_{swap}$ and the fidelity after swapping is \cite{Munro15} \begin{equation}
F_{swap} = F_1F_2 + (1-F_1)(1-F_2)/3\label{eq:F_swap}
\end{equation} where $F_1$ and $F_2$ are the fidelities of the two original pairs. 

\section{Parallel Swapping}\label{part:parallel_swapping}
\subsection{The Blind Swapping protocol}

\begin{figure}[hbtp]
	\centering
	\includegraphics[width=0.4\textwidth]{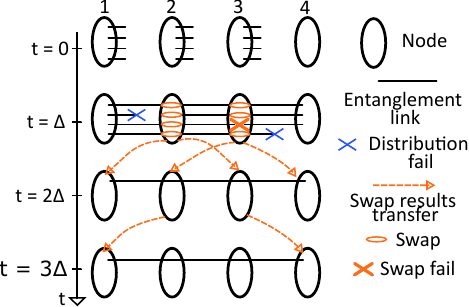}
	\caption{Scheme for the BS distribution protocol with length $l_r = 4$ and $n = 4$ generations per node}
	\label{fig:time_consumption_scheme_swap_asap}
\end{figure}

The Blind Swapping (BS) protocol is the most straightforward one and considers that $n$ pairs are generated in parallel at each node, distributed and then swapped. This protocol is illustrated in Fig.~\ref{fig:time_consumption_scheme_swap_asap} for $n = 4$ for each node and a route of $l_r = 4$ nodes. The results of all swaps are progressively sent to the destination node, so that it can then compute the state of the final end-to-end pairs.

\subsubsection{Success probability}

Let $\llbracket 1, l_r-1\rrbracket\stackrel{def}{=}\{1, \ldots, l_r-1\}$. For $i \in \llbracket 1, l_r-1\rrbracket$, let $$p_i = e^{-L_i(\alpha_i - \frac{1}{c\tau_{i+1}})-\Delta(\frac{1}{\tau_i}+\frac{1}{\tau_{i+1}})}$$ be the probability for a pair to be distributed from node $i$ to its neighbor $i+1$. The probability for one generation at each source and one swap at each node to create an end-to-end pair is
\begin{equation}
	q_{BS} = p_{swap}^{l_r-2}\prod^{l_r-1}_{i=1}p_i\label{eq:q}
\end{equation} for $l_r$ nodes, because of events independence. Let us consider that each generation is identical with the same individual success probability for distribution and swap. Then the general success probability follows a binomial law $\mathcal{B}(n, q_{BS})$ \cite{Shi19routing}. 

The event corresponding to a successfully established end-to-end pair is denoted by $e2e$. The probability to have at least one pair is thus 
\begin{equation}
\P(e2e \geq 1) = 1-(1-q_{BS})^n.\label{eq:Pe2e_geq_1_bs}
\end{equation}

The time needed for distributing all pairs is $\Delta$, as shown on Fig. \ref{fig:time_consumption_scheme_swap_asap}. We make the assumption that swapping is immediate, so one more time step will be needed to transfer the swap results of each node to its neighbor. As such, it takes a time $(l_r-2)\times\Delta$ to transfer all swap results to the destination node of the route. The protocol total execution time is $T_{BS} = (l_r-1)\Delta.$

\subsubsection{Resources consumption} We propose to evaluate the resource consumption as the minimum amount of generated pairs needed to establish at least one end-to-end pair on a route. The link probability is greater than $1-\varepsilon $ for some $\varepsilon \in ]0,1[$ if and only if the number of generated pairs at each node is greater than 
\begin{equation}
n^*_{BS} = \Big\lceil\frac{\ln(\epsilon)}{\ln(1 - q_{BS})}\Big\rceil,
\label{eq:n_min_bs}
\end{equation} so BS consumes $n_{BS} = n^*_{BS}(l_r-1)$.

\subsubsection{Link fidelity}

The fidelity drops for all pairs while they travel to the nearest node, during the successive swaps, and finally during the waiting time to transfer the swap results. As such we may define a function $h_{\tau}:F\mapsto F_{swap}(F,F_{\tau_{coh}}(\Delta, 1))$ so that the link fidelity for the BS protocol is \begin{equation}
    F_{BS} = F_{\tau}((l_r-2)\Delta,\underset{i=2}{\overset{l_r-2}{\circ}}h_{\tau_{coh}}(F_{\tau_{coh}}(\Delta,1)))
\end{equation} where $\tau = \min(\tau_1,\tau_{l_r})$. We take $\tau_{coh}$ as the equivalent of the coherence time in the fiber.

\subsection{The Heralded Distribution protocol}

\begin{figure}[hbtp]
	\centering
	\includegraphics[width=0.45\textwidth]{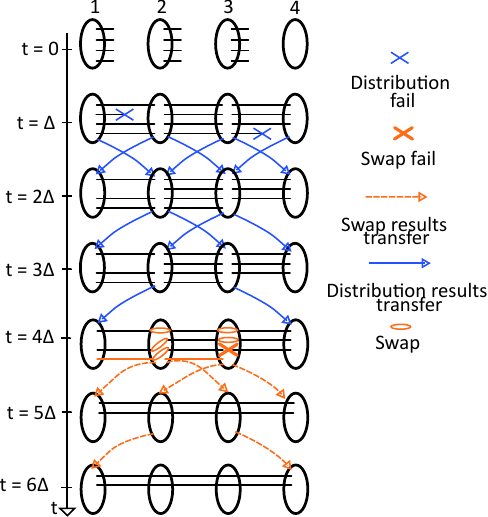}
	\caption{Scheme for the HD distribution protocol with length $l_r = 4$ and $n = 4$ generations}
	\label{fig:time_consumption_scheme_abane}
\end{figure}

The second parallel protocol, referred to as Heralded Distribution (HD) \cite{Abane25}, increases the success probability by only swapping fully distributed pairs together. Fig.~\ref{fig:time_consumption_scheme_abane} shows the process in which $(l_r-2)$ time slots are used for communication between neighbors. The transmitted data inform the nodes of lost photons, so that they will not swap lost pairs. During this time, all pairs are stored in the node memories. The total duration of HD is $T_{HD} = 2(l_r-1)\Delta$.

\subsubsection{Success probability}
This protocol makes $n$ generations but swaps only the minimum number of pairs in common with all nodes. Let $M^n_{l_r}$ be the minimum common number of distributed pairs on a route of length $l_r$. The end-to-end probability $e2e |M_{l_r}^n=m$ follows a binomial law $\mathcal{B}(q_{HD},m)$ with
\begin{equation}
q_{HD} = p_{swap}^{l_r-2}e^{-(l_r-2)\Delta(\frac{1}{\tau_1}+\frac{1}{\tau_{l_r}})}e^{-\Delta(l_r-1)\sum_{i = 1}^{l_r-1}(\frac{1}{\tau_i}+\frac{1}{\tau_{i+1}})},
\end{equation} where the first exponential describes the communication of the swap results and the second one describes the communication of the distribution results before swapping. 

Following the total law of probabilities, the general probability to have exactly $k$ end-to-end pairs writes for  $ k\in\llbracket 0, n\rrbracket$ \cite{Abane25},
\begin{equation}
\P(e2e = k) = \sum_{m = k}^{n}\P(e2e = k | M^n_{l_r} = m)\P(M^n_{l_r} = m).
\end{equation}
The expression of $M^n_{l_r}$ follows a recursion proposed by \cite{Abane25}. If $S_i$ is the number of distributed pairs from node $i$ and follows a binomial law $\mathcal{B}(n,p_i)$, then we may analytically write that 
\begin{equation}
		\P(M^n_{l_r} = m) = \P(M^n_{l_r} \geq m) - \P(M^n_{l_r} > m)
			\label{eq:P_min}
\end{equation}
with $$\P(M^n_{l_r} > m) = \prod_{i = 1}^{l_r-2}\P(S_i > m).$$
Hence,
\begin{multline}
\label{pe2e} \P(e2e >0) = \sum_{m = 1}^{n} (1-(1-q_{HD})^{m}) \\\left(\prod_{i = 1}^{l_r-2}\P(S_i > m-1) - \prod_{i = 1}^{l_r-2}\P(S_i > m)  \right)  .    
\end{multline}

\subsubsection{Resources consumption}

For the establishment of at least one pair with a certainty given by an $\varepsilon\in]0,1[$, we need to generate at least $n^*_{HD}$ pairs at each node. This value is the smallest $n$ for which \begin{equation}
   1- \P(e2e >0) < \varepsilon.
\end{equation} 
where $\P(e2e >0)$  is defined by Equation~\eqref{pe2e}. Like BS, this protocol generates the same amount of pairs at each node except for the destination, so that the total number of consumed pairs is $n_{HD} = n^*_{HD}(l_r-1)$.

\subsubsection{Link fidelity}
The fidelity for the HD protocol is expressed by using the same arguments as for BS protocol, except for the additional waiting time of $(l_r-1)\Delta$ before swapping. The function $h_{\tau}$ becomes $\Tilde{h}_{\tau}:F\mapsto F_{swap}(F,F_{\tau_{coh}}(l_r\Delta, 1))$. The link fidelity for the HD protocol is then \begin{equation}
    F_{BS} = F_{\tau}((l_r-2)\Delta,\underset{i=2}{\overset{l_r-2}{\circ}}\Tilde{h}_{\tau_{coh}}(F_{\tau_{coh}}(\Delta,1))).\label{eq:F_hd}
\end{equation}

\section{Sequential Distribution}\label{part:sequential_distribution}
\subsection{The Progressive Swapping protocol}

\begin{figure}[hbtp]
	\centering
	\includegraphics[width=0.45\textwidth]{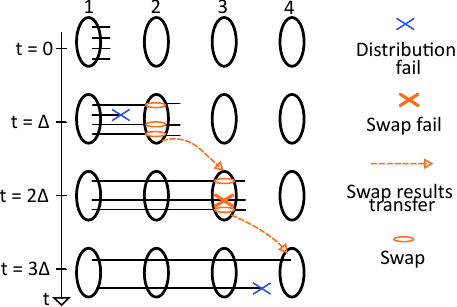}
	\caption{Scheme for the PS with route length $l_r = 4$ and $n = 4$ generations}
	\label{fig:time_consumption_scheme_loss_model}
\end{figure}

The Progressive Swapping (PS) \cite{Wang24_SES_noF_nor_mem} is illustrated Fig.~\ref{fig:time_consumption_scheme_loss_model} and, unlike the two others, progresses linearly along the path. This has the advantage of swapping the received pairs with fresh ones at each node for a better fidelity and less resource consumption. As shown, $n$ pairs are generated at the source node, then distributed and stored in the second node. The second node counts how many photons have been received and only generates and swaps this amount of photons. The results of the swaps are transmitted along the path at the same time as the qubits.

\subsubsection{Success probability}

Let us denote by $R_i$ the number of received photons and $E_i$ that of emitted photons (successful swaps) at node $i$. The source node generates and sends $n$ photons. Each node $i$ receives a photon transmitted from the previous node with probability $p_{i-1}$, and emits a swapped photon with probability $b_i$, the success probability of swapping at node $i$. 

When repeated, these become binomial processes and we can hence define two sequences of conditional events:
\begin{align}
	&E_i | R_i \stackrel{d}{=} \mathcal{B}(R_i, b_i)\\
	&R_i | E_{i-1}\stackrel{d}{=} \mathcal{B}(E_{i-1}, p_{i-1})
\end{align}
where $\stackrel{d}{=}$ denotes equality in distribution. 
By definition, there are $R_{l_r}$ end-to-end pairs. The probability law of $R_{l_r}$ can be computed out of its generating function and it follows that for $k\in\llbracket 2, l_r\rrbracket$,
\begin{equation}
	\E_k(z^{R_k}) = (1 - (1 - z)q_{PS,k})^{n} \text{ with } q_{PS,k} = q_1\prod_{j=2}^{k-1}b_jq_j\label{eq:q_ps_k}
\end{equation} where $$\forall j \in\llbracket 1,l_r-1\rrbracket,~ q_j = e^{-L_i(\alpha_i - \frac{1}{c\tau_{i+1}})-\Delta(\frac{i}{\tau_1}+\frac{1}{\tau_{i+1}})}.$$
This generating function is that of a binomial law so that for all $k\in\llbracket 2,l_r\rrbracket$, $$R_k \stackrel{d}{=}{B}\left(n, q_1\prod_{j=2}^{k-1}b_jq_j\right).$$ 

Since the distribution happens node by node, the swap results are sent at the same time as the swapped photons. Hence, the total duration is $T_{PS} = (l_r-1)\Delta$. 

\subsubsection{Resources consumption}
The success probability of at least one pair is greater than $1-\varepsilon $ for some $\varepsilon \in ]0,1[$ if and only if the number of generated pairs at the source node is greater than 
\begin{equation}
n^*_{PS} = \Big\lceil\frac{\ln(\epsilon)}{\ln(1 - q_{PS,l_r})}\Big\rceil.
\label{eq:n_min_ps}
\end{equation}

This protocol only generates as many pairs as have been received at intermediary nodes. We may compute the average of this value, knowing that $\langle R_i \rangle = n^*_{PS}q_{PS,i}$ so that the mean number of consumed resources is
\begin{equation}
n_{PS} = n^*_{PS}\left(1+\sum_{i = 2}^{l_r-1}q_{PS,i}\right).
\end{equation}

\subsubsection{Link fidelity}

The fidelity of this protocols follows a series of distributions with coherence time $\tau_{coh}$, and swaps with new pairs of fidelity $1$. We may define a function $g_{\tau_{coh}}:F\mapsto F_{\tau_{coh}}(\Delta,F_{swap}(1,F))$ so that \begin{equation}
F_{PS} = \underset{i=1}{\overset{l_r-2}{\circ}}g_{\tau}(F_{\tau_{coh}}(\Delta,1)).\label{eq:F_ps}
\end{equation}

\subsection{Progressive Swapping to the Middle}

\begin{figure}[hbtp]
	\centering
	\includegraphics[width=0.45\textwidth]{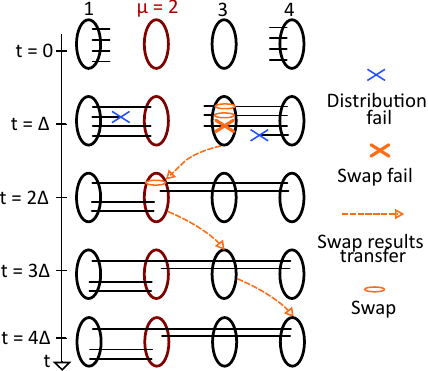}
	\caption{Scheme for the progressive swapping in the middle protocol with $l_r = 4$ nodes and $n = 4$ generated pairs, the chosen middle is $\mu = 2$}
	\label{fig:scheme_psm}
\end{figure}

We now propose a distribution protocol that takes advantage of the sequential resource sparing of PS and of the high efficiency of the parallel protocols. PSM uses two PS distributions meeting at a middle node $\mu$, where they swap. The swap results all converge to node $\mu$ before traveling to the destination node $l_r$, as shown Fig. \ref{fig:scheme_psm}. If $D_{\mu} = l_r-\mu$ is the number of links from $\mu$ to $l_r$ and $G_{\mu} = \mu - 1$ from $1$ to $\mu$, then this protocols takes a total time equal to $2\max(G_{\mu},D_{\mu})\Delta$. We usually choose $\mu = \lfloor \frac{l_r}{2}\rfloor$, so that the total time is \begin{equation}
    T_{PSM} = 2\left(l_r- \Big\lfloor\frac{l_r}{2}\Big\rfloor\right)\Delta.
\end{equation}

\subsubsection{Success probability}
The success probability follows \begin{equation}\begin{split}
   & \P(e2e = k) = \sum_{m = 0}^{n}\P(e2e = k | \min(G_{\mu},D_{\mu}) = m)\\
   &\times \P(\min(G_{\mu},D_{\mu}) = m),
\end{split}
\end{equation} with $e2e|\min(G_{\mu},D_{\mu}) \stackrel{d}{=}\mathcal{B}(\min(G_{\mu},D_{\mu}), b_{\mu}e^{-d\Delta(\frac{1}{\tau_1}+\frac{1}{\tau_{l_r}})})$. Similarly to the case for HD, 
\begin{align*}
  &  ~\P(\min(G_{\mu},D_{\mu}) = m) \\
    = & ~\P(\min(G_{\mu},D_{\mu}) \geq m) - \P(\min(G_{\mu},D_{\mu}) > m)
\end{align*}
with $\P(\min(G_{\mu},D_{\mu}) > m) = \P(G_{\mu} > m)\P(D_{\mu} > m)$.

The left and right probabilities follow binomial laws so that
\begin{align}
  & G_{\mu} \stackrel{d}{=}\mathcal{B}(n,q_{PS,\mu}) \text{ from \eqref{eq:q_ps_k} and }\\
  & D_{\mu} \stackrel{d}{=} \mathcal{B}(n,d_{l_r}\prod_{i = \mu +1}^{l_r-1}b_id_i)
\end{align}where $d_i = e^{-L_i(\alpha_i - \frac{1}{c\tau_{i}})-\Delta(\frac{l_r-i}{\tau_{l_r}}+\frac{1}{\tau_{i}})}$.

\subsubsection{Resources consumption}
In order to build at least one pair with a certainty given by an $\varepsilon\in]0,1[$, we need to generate at least $n^*_{PSM}$ pairs at the source node. This value is the smallest $n$ so that \begin{equation}
    \sum_{m = 0}^{n}\P(e2e = 0 |m)\P(m) < \varepsilon.
\end{equation} Similarly to PS, this protocol consumes the number of pairs received at each intermediary node from $2$ to $\mu-1$ and from $l_r-1$ to $\mu+1$ so that the mean number of consumed resources is \begin{equation}
n_{PSM} = n^*_{PSM}\left(2 + \sum_{i = 2}^{\mu-1}q_{PS,i}+\sum_{i = \mu+1}^{l_r-1}d_{l_r}\prod_{j = i}^{l_r-1}b_jd_j\right).
\end{equation}

\subsubsection{Link fidelity}

In this case, the fidelity of the pairs arriving from the right and from the left follow the law defined for PS for chains of $G_{\mu}$ and $D_{\mu}$ links respectively, before being swapped once at node $\mu$ and then waiting at the extremities for a duration $D_{\mu}\Delta$. Thus for $\tau = \min(\tau_1,\tau_{l_r})$,\begin{multline}
 F_{PSM} = F_{\tau}(D_{\mu}\Delta,F_{swap}(F_{PS}(l_r = G_{\mu}+1),\\
 F_{PS}(l_r = D_{\mu}+1))).\label{eq:F_psm}\end{multline}

\section{Performance}\label{part:performances}

\begin{figure}[htpb]
    \centering
    \includegraphics[width=1\linewidth]{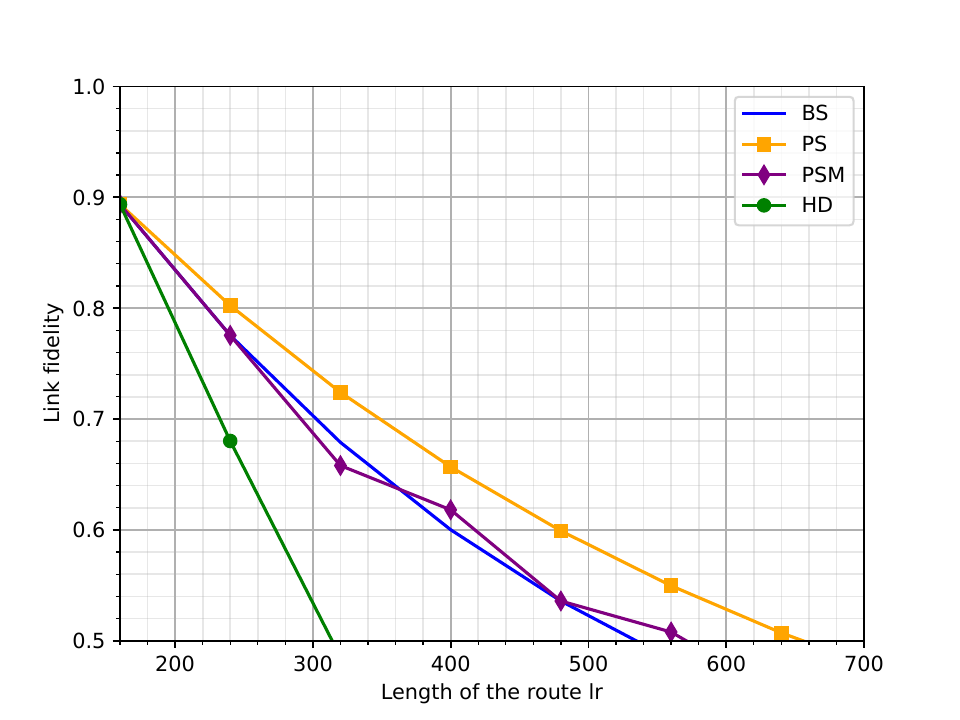}
    \caption{Link fidelity as a function of the length of the route for a distance of $L = 80km$ between nodes, $n = 30$ generations, a memory life-time $\tau = 40ms$, a time-slot of $\Delta = 2ms$ and a swap probability of $p_{swap} = 0.9$}
    \label{fig:F}
\end{figure}

\begin{figure}[htpb]
    \centering
    \includegraphics[width=1\linewidth]{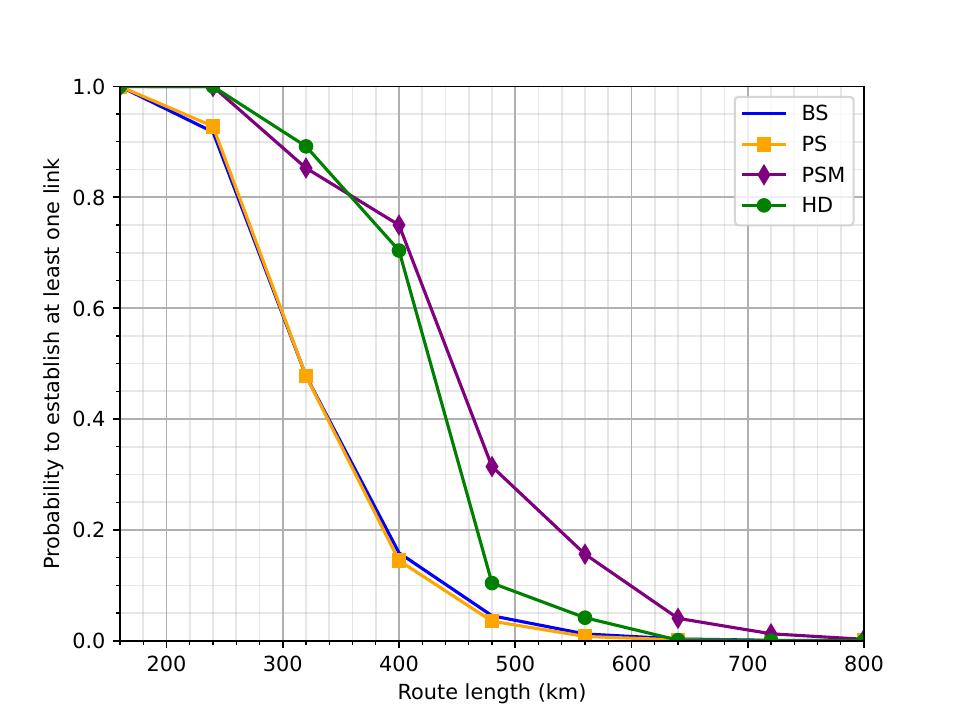}
    \caption{Probability to establish at least one end-to-end pair as a function of the length of the route for a distance of $L = 80km$ between nodes, $n = 30$ generations, a memory life-time $\tau = 40ms$, a time-slot of $\Delta = 2ms$ and a swap probability of $p_{swap} = 0.9$}
    \label{fig:P}
\end{figure}

\begin{figure}[htpb]
    \centering
    \includegraphics[width=1\linewidth]{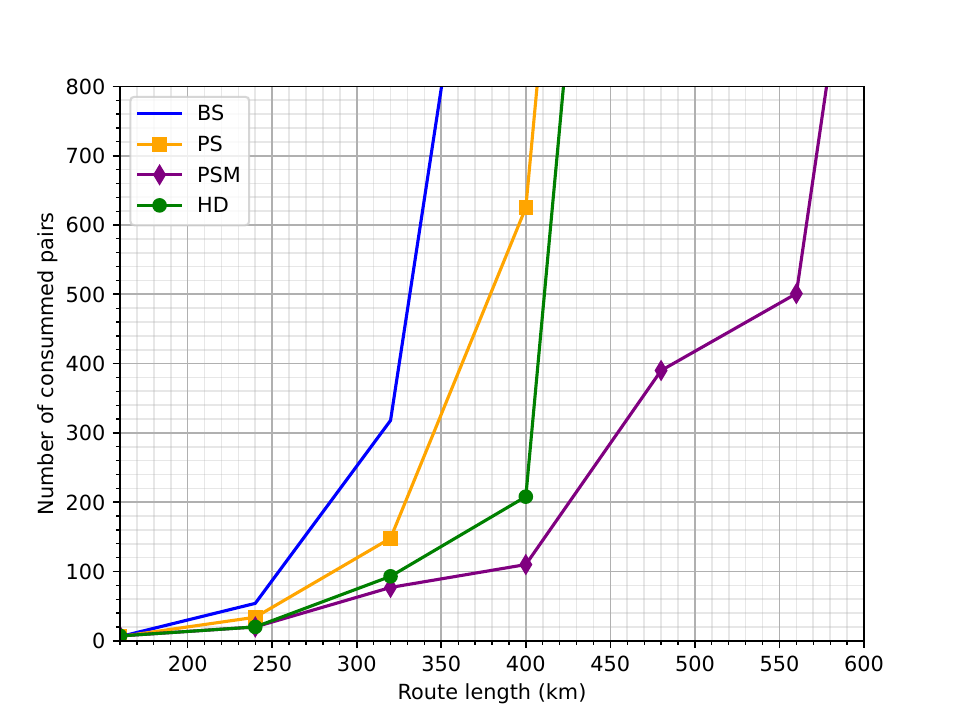}
    \caption{Total number of generated pairs during each protocol as a function of the length of the route for a distance of $L = 80km$ between nodes, $n = 30$ generations, a memory life-time $\tau = 40ms$, a time-slot of $\Delta = 2ms$ and a swap probability of $p_{swap} = 0.9$}
    \label{fig:n_conso}
\end{figure}

We compare the  four protocols in terms of link fidelity, link probability, and total resource consumption. The results are displayed Fig.\ref{fig:F}, Fig. \ref{fig:P}, and Fig. \ref{fig:n_conso} respectively.

We observe the performances for imperfect quantum memories with a relatively average memory life-time $40ms$. Quantum memories offer storage for durations from a few microseconds (superconducting technologies \cite{Bland25_transmon_QM_1ms}) to a few seconds (atomic qubits \cite{Manetsch_2025_QM_12s}). In practice, we want a multimodal memory to store many states at once. At best they can store a few hundred states at once for a life-time of a few hundred $ms$ \cite{Zhang_23_QM_330modes_500ms}.

In Fig. \ref{fig:F} we compare the link fidelity of the protocols. It is an important parameter for routing qualitative quantum information, and sequential protocols achieve better values than parallel ones. We observe that it is mostly true for PSM, as it is about as good as BS. The oscillating values are due to the dependency of the fidelity in the parity of the number of nodes.
Fidelity requirements are strict for any network application and Fig. \ref{fig:F} shows that the distribution protocols deliver low values. We consider perfect sources, but in practice the initial fidelity is not quite $1$, which lowers the link fidelity some more. Purification protocols are used to create higher fidelity pairs out of average fidelity ones. These protocols require a minimum fidelity of $F = 0.5$ and their efficiency decreases for links with higher fidelity. They are especially greedy in link resources and can be scheduled in the distribution protocol \cite{Victoria23_pur_UMASS}, or in the routing protocol \cite{li2020_C0_ent_pur}. Purification is beyond the scope of this work.

\begin{figure}[htpb]
    \centering
    \includegraphics[width=1\linewidth]{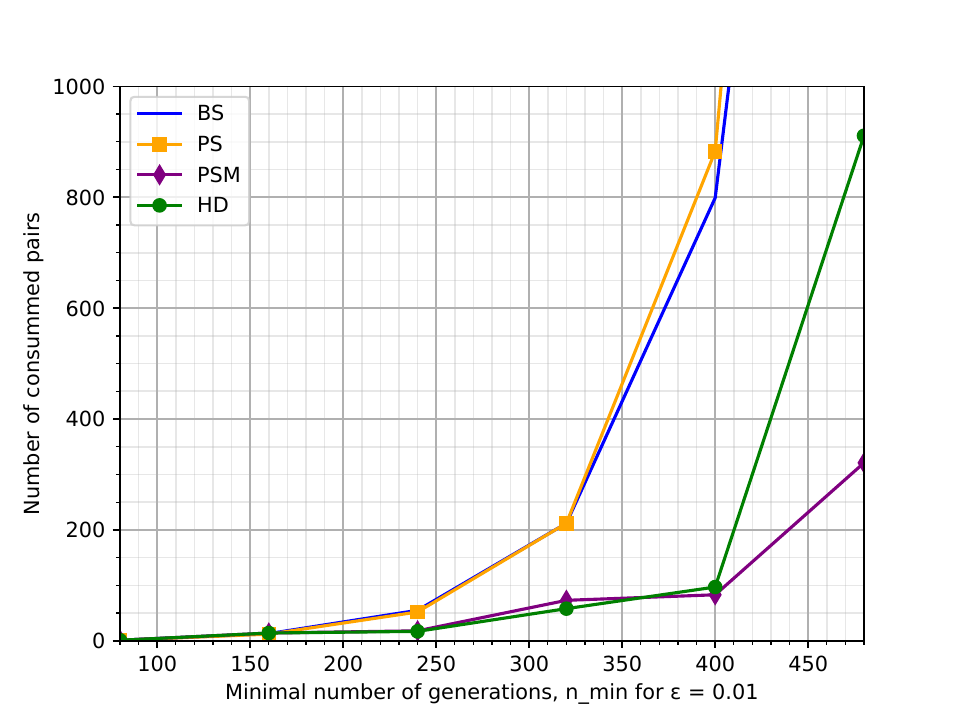}
    \caption{Minimal number $n^*$ of generations per node or at the start node as a function of the length of the route}
    \label{fig:nmin}
\end{figure}

In Fig. \ref{fig:P} we observe the success probability of the protocols as a function of the length of the route with a repeater node every $80km$. It shows the advantage of our PMS protocol even compared to the parallel HD. The Fig. \ref{fig:nmin} completes the link probability performance showing the minimal amount of pairs to be generated at all nodes or at the start node, depending on the protocol. It mirrors Fig. \ref{fig:P} with PSM being about as efficient as HD, and even better for longer distances. It is worth noting that for better memories the HD protocol can be more efficient than all the others, especially as distances increase. All parameters are the same in all figures.

A good success probability lowers the requirement in terms of  the number of pairs to be initially generated, so that it optimizes the total consumption. As shown on Fig. \ref{fig:n_conso}, combined with the inherent resource economy of the sequential protocols, PMS high success probability makes it the most economical protocol in terms of generated pairs. It is still pretty close to HD, especially when considering better memories and shorter distances. Its advantage can however still be harnessed through loss model centered routing such as in \cite{Wang24_SES_noF_nor_mem} or \cite{Yang25_routing_best_resource_alloc_ses}.

\section{Conclusion}
\label{part:conclusion}

In this paper we have developed the Progressive Swapping to the Middle (PSM) protocol that takes advantage of sequential protocol properties to save resources, with an efficiency in success probability comparable to  that of the best parallel protocols. It is a reasonably fast protocol that guarantees a good end-to-end entangled pair  fidelity. 

We observe an overall advantage in resource consumption for PSM in the case of short lived quantum memories and repeaters placed at less than $100km$ from each other. Better memories would allow for a more efficient parallel protocol. In a next step, we shall consider the problem of designing a routing protocol that exploits the performance of PSM, similarly to what has been done for the PS protocol.

\section*{Acknowledgments}

We acknowledge the support of the European Union’s Horizon Europe research and innovation program through the Quantum Secure Network Partnership project (QSNP).

\balance

\bibliographystyle{IEEEtran}
\bibliography{ref}

\end{document}